\documentclass[12pt]{iopart}
\usepackage{iopams}  
\usepackage{bm}
\usepackage{graphicx}
\usepackage{xcolor}
\usepackage{cite}
\usepackage[colorlinks=true,linkcolor=blue,citecolor=blue,urlcolor=blue]{hyperref}
\usepackage[all]{hypcap} 

\newcommand{\dd}[0]{\mathrm{d}}
\newcommand{\bfsog}{\mbox{\boldmath${\cal G}$}}
\newcommand{\bfsok}{\tilde{\mbox{\boldmath${\cal G}$}}}
\newcommand{\bfsox}{\mbox{\boldmath${\cal X}$}}
\begin{document}
\title[Statistics of work performed by optical tweezers]{Statistics of work performed by optical tweezers with general time-variation of their stiffness}

\author{Petr Chvosta$^1$, Dominik Lips$^2$, Viktor Holubec$^{1,3}$, \\ Artem Ryabov$^1$, and Philipp Maass$^2$}

\address{$^1$ Charles University, 
Faculty of Mathematics and Physics, 
Department of Macromolecular Physics, 
V~Hole\v{s}ovi\v{c}k\'{a}ch~2, 
CZ-180~00~Praha, Czech Republic}
\address{$^2$ Universit\"{a}t Osnabr\"{u}ck, 
Fachbereich Physik, 
Barbarastra{\ss}e~7, 
49076 Osnabr\"{u}ck, Germany}
\address{$^3$ Universit{\"a}t Leipzig, 
Institut f{\"u}r Theoretische Physik, 
Postfach 100~920, 
D-04009 Leipzig, Germany}

\eads{\mailto{Petr.Chvosta@mff.cuni.cz}, \mailto{dlips@uos.de}, \mailto{viktor.holubec@gmail.com}, \mailto{maass@uos.de}, \mailto{rjabov.a@gmail.com}}
\vspace{10pt}
\begin{indented}
\item[] March 2020
\end{indented}

\begin{abstract}
We derive an exact expression for the probability density of work done on a particle that diffuses in a parabolic potential with a stiffness varying by an arbitrary piecewise constant protocol. Based on this result, the work distribution for time-continuous protocols of the stiffness can be determined up to any degree of accuracy. This is achieved by replacing the continuous driving by a piecewise constant one with a number $n$ of positive or negative steps of increasing or decreasing stiffness. With increasing $n$, the work distributions for the piecewise protocols approach that for the continuous protocol. The moment generating function of the work is given by the inverse square root of a polynomial of degree $n$, whose coefficients are efficiently calculated from a recurrence relation. The roots of the polynomials are real and positive (negative) steps of the protocol are associated with negative (positive) roots. Using these properties the inverse Laplace transform of the moment generating function is carried out explicitly. Fluctuation theorems are used to derive further properties of the polynomials and their roots. 
\end{abstract}

\vspace{2pc}

\submitto{\JPA}
\maketitle

\section{Introduction}
Work, heat and entropy, when defined microscopically for individual
trajectories have attracted much attention in nonequilibrium
statistical physics \cite{Sekimoto:1997, Seifert:2005, Ritort:2008, Esposito/etal:2009}, and form
key quantities in stochastic thermodynamics
\cite{Seifert:2012}. Probability distributions of work 
fulfil detailed and integral fluctuation theorems \cite{Crooks:2000, Jarzynski:2008, Harris/Schuetz:2007,    Esposito/vandenBroeck:2010, VandenBroeck/Esposito:2010, Jarzynski:2011, Bochkov/Kuzovlev:2013, Maillet/etal:2019}. These apply to nonequilibrium systems that
are externally driven by a time-dependent protocol of control
variables. The theorems hold true universally and allow one to
determine free energy differences between equilibrium states from nonequilibrium processes. They represent refinements of the second law of thermodynamics and can be applied to extract equilibrium information from nonequilibrium measurements \cite{Hummer:2002, Shirts/etal:2003, Collin/etal:2005, Pohorille/etal:2010, Kim/etal:2012, Dellago/Hummer:2014,Ribezzi2014, Halpern/Jarzynski:2016, Arrar/etal:2019} either by using the Crooks theorem~\cite{Crooks:1998, Crooks:1999} or the Jarzynski identity~\cite{Jarzynski:1997, JarzynskiPRE:1997}. For the Crooks theorem one needs to compare work distributions for a time-forward and a time-reversed process \cite{Severino/etal:2019}, which is not always possible in practice. A direct application of the Jarzynski identity would require to extend measured work probability distributions to large negative values. For this reason, it is important to have analytical expressions for work distributions for most common situations~\cite{Palassini2011,Holubec/etal:2015}.

Since forces applied in single-molecule experiments are often linear with respect to particle displacements, as, for example, those exerted by optical tweezers \cite{Trepagnier/etal:2004, Carberry/etal:2004, Carberry/etal:2007, Andrieux/etal:2007, Khan/Sood:2011, Mestres/etal:2014}, the consideration of Brownian motion in time-varying harmonic potentials is particularly relevant. For the variation in time, two kinds of operation can be distinguished: in the first the minimum of the harmonic potential is changed with time (``moving parabola'') and in the second the stiffness of the potential is varied (``breathing parabola''). Analytical results for the work distribution could be derived for specific protocols in both modes of operation. Moreover, theories were developed to predict the exact asymptotic behaviour of work distributions \cite{Engel:2009, Nickelsen/Engel:2011, Noh/etal:PRL2013, Ryabov/etal:2013, Holubec/etal:2015, Manikandan/Krishnamurthy:2017}. For the moving parabola with a potential minimum changing with time, the work distribution was calculated in Refs.~\cite{vanZon/Cohen:2003, vanZon/Cohen:2004, Cohen:2008, Nickelsen/Engel:2011, Subasi/Jarzynski:2013, Kim/etal:2014}.  
For general protocols of the breathing parabola, the problem can be reduced to the task of solving the Riccati equation \cite{Ryabov/etal:2013}, or, equivalently, of solving a system of coupled ordinary differential equations \cite{Speck:2011,Kwon/etal:2011, Kwon/etal:2013}. Explicit analytic results could be obtained for specific protocols only: (i) for a single sudden change \cite{Kwon/etal:2011}, (ii) for slow driving, where the work distribution
is Gaussian \cite{Speck:2011, Hoppenau/Engel:2013}, and (iii) and for a particular rational stiffness function \cite{Ryabov/etal:2013, Kwon/etal:2013}. An Onsager-Machlup type of theory has been developed
to obtain approximate solutions \cite{Deza/etal:2009}.

In the breathing parabola problem, the derivation of analytical solutions for work distributions seems to be impossible for general protocols. Here we derive an explicit solution for the joint probability of work and particle position, which is exact for protocols consisting of a sequence of instantaneous changes (steps) of the stiffness. This solution can be used to determine work distributions for time-continuous protocols with arbitrary accuracy by systematically increasing the number of steps in a discretization of the respective protocol. In contrast to general algorithms based on a numerical solution of the Fokker-Planck equation~\cite{Holubec2019}, the present methodology significantly reduces computational costs. Compared to purely numerical schemes \cite{Paolella:2018}, it provides valuable analytical insights. Besides the approximation of continuous protocols, our exact results can be directly applied to a possible experiment with piecewise constant values of experimental parameters such as laser power and temperature. 

The paper is structured as follows. In Sec.~\ref{sec:recipe} we outline
the general method to obtain work distributions based on a few key equations, and
demonstrate its power for a representative example. In Sec.~\ref{sec:laplace-proof} 
we present the methodology of our approach. The main results are explicit formulas for
the joint probability of work and particle position [Eqs.~(\ref{Knsolution}) and (\ref{Rnsolution})], 
and for the moment generating function of the work [Eqs.~(\ref{Phinufinal}) and (\ref{eq:recurrence})]
in terms of a polynomial $Q_n$. 
In  Sec.~\ref{sec:work-pdf} we derive recurrence relations for the $Q_n$, discuss properties of the roots of the polynomial and perform the inverse Laplace transform of the moment generating function.
Section~\ref{sec:summary} summarizes our main findings and discusses their broader relevance, 
including perspectives for future work.

\section{Method and representative examples}
\label{sec:recipe}
We consider the overdamped Brownian motion of a single particle in an
external harmonic potential 
\begin{equation}
V(x,t)=\frac{1}{2} k(t) x^2\,,
\label{potential}
\end{equation}
with time-varying stiffness $k(t)$. The stochastic process of
the particle position is denoted by $\mathsf{X}(t)$ and evolves in time according to
the Langevin equation 
\begin{equation}
\frac{\dd\mathsf{X}(t)}{\dd t}=-\mu k(t) \mathsf{X}(t)+\sqrt{2D}{\boldsymbol\zeta}(t)\,,
\end{equation}
where $D$ is the diffusion constant, $\mu=\beta D$ the mobility,
and $\beta=1/(k_{\rm B}T)$, with $k_{\rm B}$ the Boltzmann constant
and $T$ the temperature; $\boldsymbol\zeta(t)$ is a Gaussian white noise with zero mean and correlator $\left<\boldsymbol\zeta(t)\boldsymbol\zeta(t') \right> = \delta(t-t')$.

Each stochastic trajectory
in a time interval $[0, t]$ yields a value of the work done on the
particle, i.e.\ the work process $\mathsf{W}(t)$ is a functional of
the Markov process $\mathsf{X}(t')$, $0 \leq t' \leq t$. 
It can be defined based on the first law 
${\rm d}V = (\partial_x V){\rm d}x + (\partial_t V){\rm d}t = \delta Q + \delta W$, 
as the change of the particle energy related to the time-variation of the external potential \cite{Sekimoto:1997, Seifert:2012}. 
Hence in our case, the work is a quadratic functional of the process $\mathsf{X}(t')$,
\begin{equation}
\mathsf{W}(t)=\int_0^t\dd t'\,\frac{\partial V(\mathsf{X}(t'),t')}{\partial t'}
=\frac{1}{2}\int_0^t\dd t'\,\dot k(t')\mathsf{X}^2(t')\,.
\end{equation}
 The work by itself is not a Markov process. However, the combined
process $\left(\mathsf{X}(t),\mathsf{W}(t)\right)$ is still a
time-nonhomogeneous Markov process and its joint probability density $\Phi_{\mathsf{X},\mathsf{W}}(x,w,t)$
obeys a Fokker-Planck equation [see Eq.~(\ref{FPEXW}) below].

Let $k(t)$ be a piecewise constant protocol exhibiting $n$ steps in
the time interval $[0,t]$ at time instants $t_1,\ldots,t_n$ with
$0\equiv t_0<t_{1}<t_{2}<\ldots<t_{n}<t$, $k(t) = k_j$ in each time
segment $[t_j,t_{j+1}[\,$, and starting value $k(0)=k_0>0$. In terms of the scaled stiffnesses (relaxation rates)
$\gamma_{j}=\mu k_j$, $j=0,\ldots,n$, this protocol can be written as
\begin{equation}
\label{eq:protocol}
\gamma(t)=\gamma_0+\sum_{j=1}^{n}(\gamma_{j}-\gamma_{j-1})\Theta(t-t_j)\,,
\end{equation}
where $\Theta(.)$ is the Heaviside step function [$\Theta(x)=1$ for $x\ge0$ and zero otherwise].

We are interested in the work PDF $\Phi_n(w)$ for the protocol $\gamma_n(t)$ and an initial state, where the particle is equilibrated in the potential $V(x,0)$.
 As shown in Sec.~\ref{sec:laplace-proof}, starting from the Fokker-Planck equation for
$\Phi_{\mathsf{X},\mathsf{W}}(x,w,t)$ and utilizing properties of the
propagator for the position process, one can derive an exact analytical result for the bilateral Laplace transform $\tilde\Phi_n(u)$ of $\Phi_n(w)$. The result is
\begin{equation}
\tilde\Phi_n(u)=
\int_{-\infty}^{+\infty}{\rm d}w\,{\rm e}^{-u w}\,\Phi_n(w)=\frac{1}{\sqrt{Q_{n}(u/\beta)}}\,\,
\label{eq:tildephi}
\end{equation}
with
\begin{equation}
\quad Q_{n}(\xi)=\sum_{j=0}^{n} q_j\xi^j
\label{eq:q_polynomial}
\end{equation}
a polynomial of degree $n$.
Defining 
\begin{eqnarray}
\eta_j &=& \exp \left[ -2\gamma_j(t_{j+1} - t_j) \right]\,,
\label{eq:polynomial-parameters-a}\\
\epsilon_j &=& \frac{\gamma_j - \gamma_{j-1}}{\gamma_0} \eta_{j-1} \eta_{j-2} \ldots \eta_1\,, 
\label{eq:polynomial-parameters-b}\\
\phi_j &=& \frac{\gamma_0}{\gamma_j} \frac{1-\eta_j}{\eta_j \eta_{j-1} \ldots \eta_1}\,,
\label{eq:polynomial-parameters-c}
\end{eqnarray}
the polynomial $Q_n(\xi)$ can be calculated from the recursion relation
\begin{equation}
\label{Qnrecursive}
Q_{n}(\xi)=Q_{n-1}(\xi)+\xi\epsilon_{n}
\left[1+\sum_{j=1}^{n-1} \phi_j\,Q_j(\xi)
\right]\,.
\end{equation}
The initial condition is
$Q_0(\xi)=1$ as without a step, no work is done and accordingly $\Phi_0(w, t) = \delta(w)$ and $\tilde\Phi_n(u)=1$. An explicit expression for the coefficients $q_j$ in Eq.~(\ref{eq:q_polynomial})  
is derived in Sec.~\ref{sec:polynominal-coefficients} and reads
\begin{equation}
\label{eq:qcoeefficients}
q_j=
\sum_{i_1=1}^{n-j+1}\epsilon_{i_1}\phi_{i_1-1,0}
\sum_{i_2=i_1+1}^{n-j+2}\epsilon_{i_2}\phi_{i_2-1,i_1}
\ldots
\sum_{i_j=i_{j-1}+1}^n\epsilon_{i_j}\phi_{i_j-1,i_{j-1}}\,,
\end{equation}
where
\begin{equation}
\phi_{j_2,j_1} = \sum_{j=j_1}^{j_2}\phi_j \, .
\end{equation}
Expanding both sides of Eq.\ (\ref{eq:tildephi}) into a power
series in the variable $u$, we see that the coefficients
$q_j$ determine the moments of the work $\mathsf{W}_n$ after $n$ steps. For example, the mean
work is given by $\beta\,\left\langle\mathsf{W}_n\right\rangle=q_1/2$ and the second moment by
$\beta^2\,\left\langle
\mathsf{W}_n^2\right\rangle=-q_2+3q_1^2/4$.

To invert the bilateral Laplace transform in Eq.~(\ref{eq:tildephi}), the polynomial $Q_n(\xi)$ is factored into its roots. All roots are distinct and located on the real axis.
The roots are connected with the protocol steps: each positive step [increase of stiffness, $(\gamma_{j+1}-\gamma_j)>0$] gives a negative root and each negative step [decrease of stiffness, $(\gamma_{j+1}-\gamma_j)<0$] gives a positive root. Hence the $n_-$ negative roots $\xi_j^{-} < 0$ and $n_+$ positive roots roots $\xi_j^{+} > 0$ contribute to $\Phi_n(w)$ at positive and negative $w$, respectively.  We order the $n_\pm$ roots according to their magnitude:
\begin{eqnarray}
& & 0 < |\xi^-_1| < |\xi^-_2| < \ldots < |\xi^-_{n_-}| < \infty \, , \\
& & 0< \xi^+_1 < \xi^+_2 < \ldots < \xi^+_{n_+} < \infty \, .
\end{eqnarray}
Writing 
\begin{equation}
Q_n(\xi)=q_{n}\prod_{j=1}^{n_{-}} (\xi - \xi^-_{j}) \prod_{k=1}^{n_{+}} (\xi - \xi^+_{k})\,,
\label{eq:q-factorization}
\end{equation}
the inverse Laplace transform of $\tilde\Phi_n(u)$ in Eq.~(\ref{eq:tildephi}) gives
\begin{equation}
\label{PhinwLaplace}
\Phi_{n}(w)=\frac{\beta}{\sqrt{q_n}}\int_{-i\infty}^{+i\infty}
\, \frac{\dd\xi}{2\pi i}\,
\frac{\exp(\beta w \xi)}{\sqrt{\prod_{j=1}^{n_{-}} (\xi - \xi^-_j) \prod_{k=1}^{n_{+}} (\xi - \xi^+_k)}}\, .
\end{equation}
This integral can be calculated by a contour integration along closed paths encircling the negative and
positive real axes, where branch cuts between neighbouring pairs of roots 
give definite integral contributions. The details are described in Sec.~\ref{sec:polynomial-roots}
and the final result is:
\begin{equation}
\label{eq:Phin-decomposition}
\Phi_n(w)= \Phi_n^-(w)\Theta(-w)+\Phi_n^+(w)\Theta(w) \,,
\end{equation}

\begin{equation}
\hspace{-6em}
\Phi^+_n(w)=\beta
\sum_{j=1}^{m_-}(-1)^j I_n(\beta w;\xi^-_{2j},\xi^-_{2j-1})+
\beta\left\{\begin{array}{cl}
0\,, &  n_-\;\mbox{even}\\
\hspace{-0.5em}(-1)^{1+m_-} I_n(\beta w;-\infty,\xi^-_{n_-})\,, & n_-\;\mbox{odd}
\end{array}\right.
\label{eq:phin-p}
\end{equation}
\begin{equation}
\hspace{-6em}
\Phi^-_n(w)=\beta
\sum_{j=1}^{m_+}(-1)^j I_n(\beta w;\xi^+_{2j-1},\xi^+_{2j})+
\beta\left\{\begin{array}{cl}
0\,, &  n_+\;\mbox{even}\\
\hspace{-0.5em}(-1)^{1+m_+} I_n(\beta w;\xi^+_{n_+},\infty)\,, & n_+\;\mbox{odd}
\end{array}\right.
\label{eq:phin-m}
\end{equation}
where
\begin{equation}
\label{eq:mpm}
m_\pm = \left\{\begin{array}{cl}
\displaystyle\frac{n_\pm}{2}\,, & n_\pm\; \mbox{even}\\[1em]
\displaystyle\frac{n_\pm-1}{2}\,, & n_\pm\;\mbox{odd}
\end{array}\right.
\end{equation}
and
\begin{equation}
\label{eq:In}
I_n(\theta;a,b)=C_n\int_{a}^{b}\frac{\dd x}{\pi}\,\frac{\exp(\theta x)}
{\sqrt{\prod_{j=1}^{n_-} |\xi-\xi_j^-| \prod_{k=1}^{n_+} |\xi-\xi_k^+|}}
\end{equation}
with
\begin{equation}
\label{eq:Cn}
C_n=-\frac{1}{i^{n_+}\sqrt{q_n}}=\frac{1}{
\sqrt{\prod_{j=1}^{n_-} |\xi_j^-| \prod_{k=1}^{n_+} \xi_k^+}}\,.
\end{equation}

This means that to calculate $\Phi_n(w)$, one needs to find the roots $\xi^{\pm}_j$ and to evaluate the integrals 
$I_n(\beta w;\xi^{(\pm)}_{2j},\xi^{(\pm)}_{2j+1})$ from Eq.~(\ref{eq:In}).

\begin{figure}[t]
\centering 
\includegraphics[width=0.85\textwidth]{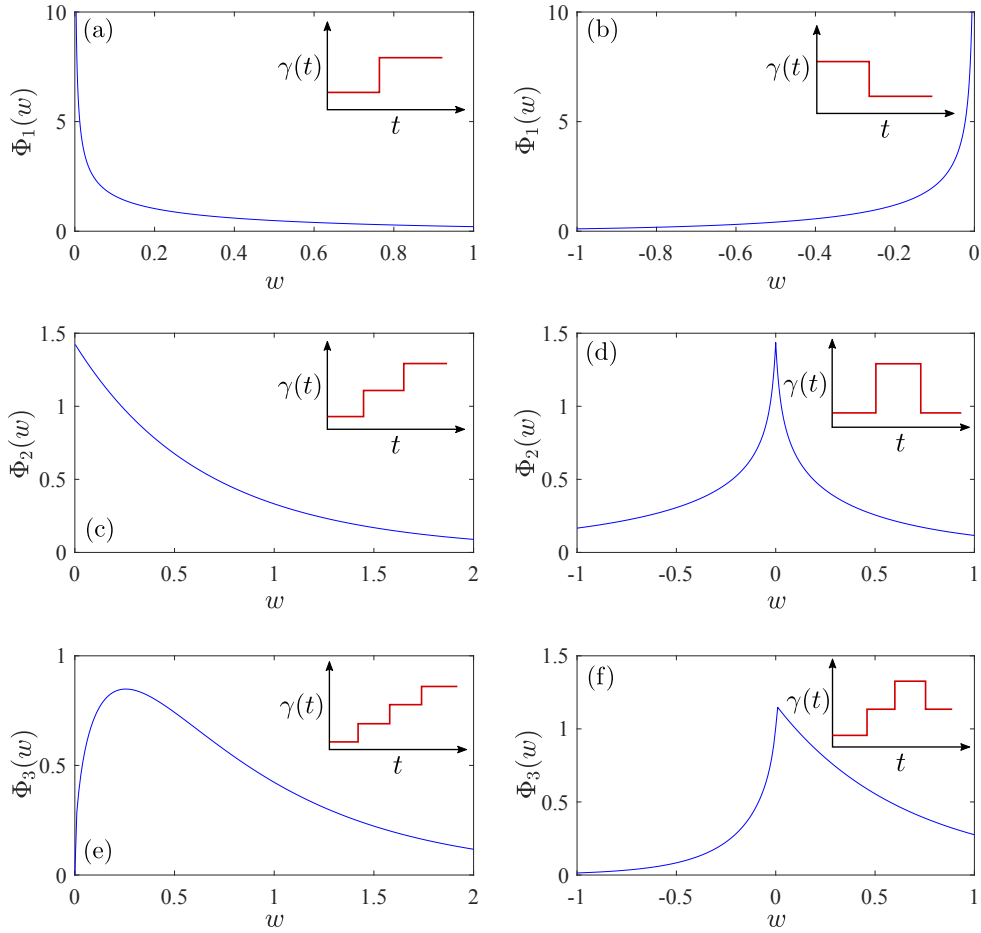}
\caption{Work PDFs $\Phi_n(w)$ for protocols with $n=1$, 2, and 3 steps, which are illustrated in the respective insets. 
In all panels except (b), the starting value of the protocol is $\gamma_0=1$. 
The duration of all protocol segments is $1$. The step size 
is always equal to $1$. In the panel (b), the time-reversed protocol of the case (a) is considered.  
Work is given in units of $k_{\rm\scriptscriptstyle B}T$, work PDFs in units of 
$\beta=1/(k_{\rm\scriptscriptstyle B}T)$, and time in units of $1/\gamma_0$.}
\label{fig:fig2-work-distributions}
\end{figure}

Figure \ref{fig:fig2-work-distributions} illustrates the work probability density function (PDF) for several protocols with a small number of steps. In all left panels [(a), (c), and (e)] the stiffness increases with time (no negative steps in the protocol). Accordingly, the work assumes positive values only. With increasing number of steps, the probability weight of small work values, corresponding to trajectories $\mathsf{X}(t)$ dwelling close to the origin during all steps, decreases as can be seen from the different behaviours of $\Phi_1$, $\Phi_2$, $\Phi_3$ around $w=0$. This can be also understood from the limit $\Phi_n(0^+) = \lim_{u \to \infty} u \tilde \Phi_n(u)$. For one step [panel (a)], we have only one root, hence the limit diverges. For two steps, there are two roots and the limit is finite [panel (c)]. For three and more steps, the limit is zero.

The right column of Fig.~\ref{fig:fig2-work-distributions} [panels (b), (d), and (e)] shows examples of work PDFs for protocols that have one negative step. Accordingly, the work can assume negative values also. The probability weight of the positive work values increases with the number of positive steps. The work PDF is always continuous at $w=0$ whenever both negative and positive steps are present in the protocol.

To demonstrate how well a work PDF for a continuous protocol can be approximated by that of a piecewise constant protocol, we consider $\gamma(t)=\gamma_{0}[2+\sin(\omega t)]$ with $\gamma_{0}/\omega=1$ in the time interval $[0, t_{\rm f}]$, $t_{\rm f}=\pi/\omega$. This corresponds to ``one sinusoidal breath of the parabola". For constructing the piecewise constant protocols, we use equidistant time steps, i.e.\ $t_j=j t_{{\rm f}}/(n+1)$, $j=0,1,\ldots,n+1$ and $\gamma_{j}=\gamma(t_j)$. The corresponding work PDF $\Phi_n(w)$ is obtained as described above: we first calculate the polynomial $Q_n(\xi)$ in Eq.~(\ref{eq:q_polynomial}) by using the recurrence relation
Eq.~(\ref{Qnrecursive}), then determine its roots $\xi^{\pm}_{j}$, and finally
evaluate Eqs.~(\ref{eq:phin-p}) and (\ref{eq:phin-m}) with the integrals given in Eq.~(\ref{eq:In}).

\begin{figure}[t!]
\centering 
\includegraphics[width=1\textwidth]{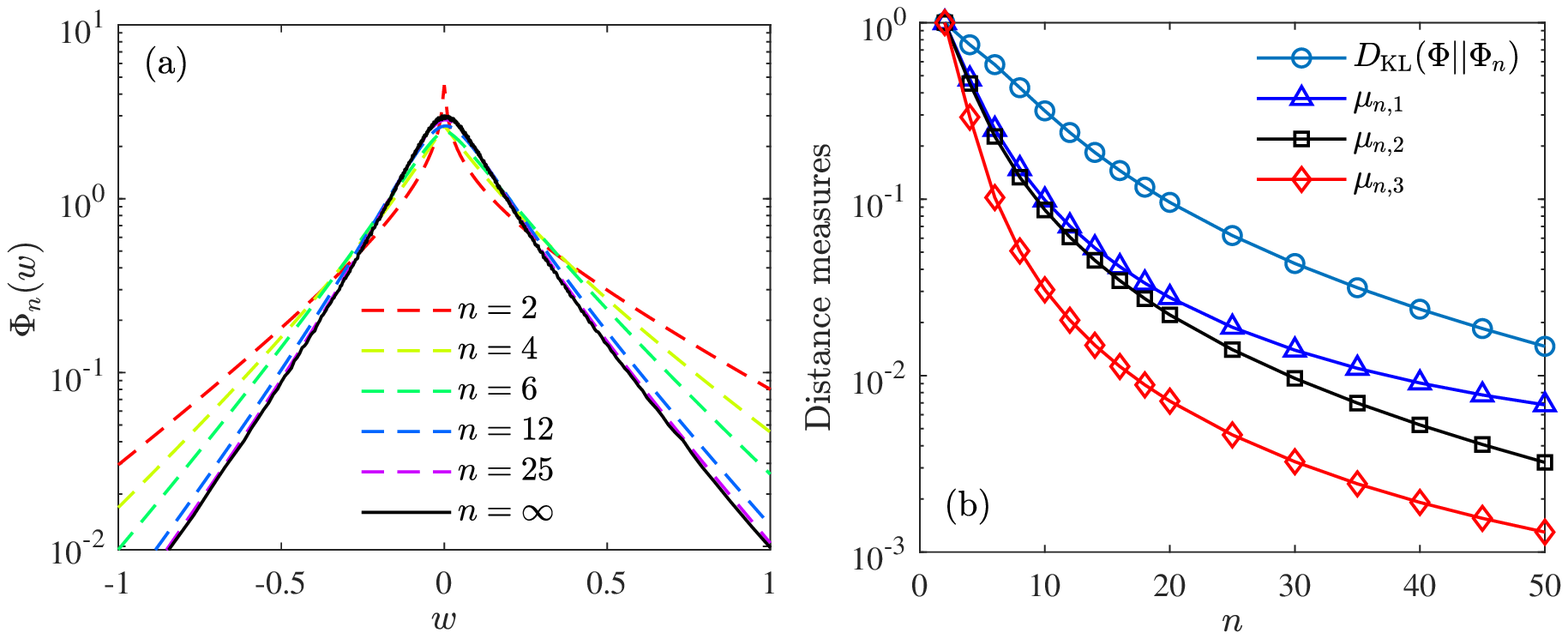}
\caption{(a) Simulated work PDF 
  $\Phi(w)$ (solid line, denoted as $n=\infty$) for the continuous protocol
  $\gamma(t)=\gamma_{0}[2+\sin(\omega t)]$ with $\gamma_0/\omega=1$
in comparison with PDFs $\Phi_n(w)$ for the piecewise constant protocols. 
(b) Kullback-Leibler divergences $D_{\rm KL}(\Phi || \Phi_n)$ of the $\Phi_n(w)$ approximations
of $\Phi(w)$ and scaled relative deviations $\mu_{n,k}$
of the $k$-th moments in the $n$-th step approximations as functions of $n$. }
\label{fig:fig1-example} 
\end{figure}

We have performed Brownian dynamics simulations for both the piecewise constant protocols and
the continuous protocol to determine the respective work PDFs based on $10^7$ realisations of the process. The simulations for the piecewise constant protocols were carried out to check our analytical results and simulation procedure, and we found perfect agreement. The simulated work PDF $\Phi(w)$ for the continuous protocol is shown in Fig.~\ref{fig:fig1-example}(a) in comparison with the analytical results for the piecewise constant protocols with different $n$. One sees how the $\Phi_n(w)$ (dashed lines) approach the $\Phi(w)$ (solid line) with increasing $n$. For just $n=12$ steps we already obtain a good approximation for $\Phi(w)$, and for $n=25$ the $\Phi_n(w)$ can hardly be distinguished from $\Phi(w)$ (for the scales of the axes used in the figure). 

To quantify the goodness of the approximation, we introduce the following distance measures:
the Kullback-Leibler divergence 
$D_{\rm KL}(\Phi || \Phi_n)=\int\dd w \, \Phi(w)\ln [\Phi(w)/\Phi_n(w)]$,
and the relative deviations 
$\Delta_{n,k}=\left|\langle \mathsf{W}^k \rangle_{\Phi_n} - \langle \mathsf{W}^k 
\rangle_{\Phi}\right|/\langle \mathsf{W}^k \rangle_{\Phi}$ of the $k$th moments in the $n$th step approximation.
The results are shown in Fig.~\ref{fig:fig1-example}(b) in a semi-logarithmic representation. To include the $\Delta_{n,k}$ in the same plot as the $D_{\rm KL}(\Phi || \Phi_n)$, 
we scaled the corresponding values with respect to $\Delta_{2,k}$, i.e.\ we plotted $\mu_{n,k} = \Delta_{n,k}/
\Delta_{2,k}$. With increasing $n$, both the $D_{\rm KL}(\Phi || \Phi_n)$ and the $\mu_{n,k}$
rapidly decrease to zero.

Let us note that it is not required to take equidistant time steps. The time discretization can be adapted specifically to a given protocol. For example, the convergence with increasing number of steps should improve by requiring the changes of the stiffness in each step to be equal. 

\section{Analytical treatment in the Laplace domain}
\label{sec:laplace-proof}

\subsection{Propagator for particle position}
\label{subsec:particle-propagator}
The particle position $\mathsf{X}(t)$ follows a time-nonhomogeneous Markov process and its
PDF $\Phi_{\mathsf{X}}(x,t)$ satisfies the Fokker-Planck equation
\begin{equation} 
\label{FPEX}
\frac{\partial}{\partial t}\Phi_{\mathsf{X}}(x,t)
=\left\{D\frac{\partial^2 }{\partial x^2}
+\mu\frac{\partial}{\partial x}
\left[\frac{\partial}{\partial x}\,V(x,t)\right]\right\} \Phi_{\mathsf{X}}(x,t)\, .
\end{equation}
For the parabolic potential (\ref{potential}) with constant stiffness, its solution is a simple Gaussian PDF, and 
for the piecewise constant protocol with $n$ jumps, the convolution of all Gaussian PDFs for the individual segments
yields again a Gaussian PDF. Here we introduce an operator notation and rescaled variances of resulting PDFs, which will be used in the derivation of the work PDF.

For an arbitrary protocol $\gamma(t)$, the Green function of Eq.\ (\ref{FPEX}) is
\begin{equation}
\label{GFX}
G(x,t|x',t')=\frac{1}{\sqrt{2\pi v(t,t')}}\,
\exp\left\{-\frac{\left[x-m(t,t')x'\right]^{2}}{2 v(t,t')}\right\}\,\,,
\end{equation}
where
\begin{eqnarray}
m(t,t')&=& \exp\left[-\int_{t'}^{t}\,\dd s\,\gamma(s)\right]\,,
\label{eq:m-v-a}\\[1ex]
v(t,t')&=& 2D\int_{t'}^{t}\,\dd s\,\exp\left[-2\int_{s}^{t}\,\dd s'\,\gamma(s')\right]\,\,.
\label{eq:m-v-b}
\end{eqnarray}

The Green function gives the solution of the Fokker-Planck equation for the
initial condition $\delta(x-x')$ at time $t'$. Notice that the
time variables $t$, $t'$ enter solely through the functions $m(t,t')$
and $v(t,t')$.  This motivates to consider the Green function as a matrix element
of an operator $\bfsog\left(m,v\right)$ in the space of continuous functions
of $x$:
\begin{equation}
\label{GFX2}
\langle x | \bfsog\left(a,b\right) | x' \rangle=\frac{1}{\sqrt{2\pi
    b}}\,\exp\left[-\frac{\left(x-ax'\right)^{2}}{2b}\right]\,
\end{equation}
for $a$, $b>0$ real.
This operator satisfies the composition rule
\begin{equation}
\label{rule1}
\bfsog\left(a_{2},b_{2}\right)\bfsog\left(a_{1},b_{1}\right)
=\bfsog\left(a_{2}a_{1},b_{2}+a_{2}^{2}b_{1}\right)\,.
\end{equation}
Similarly, if the operator
$\bfsog\left(a,b\right)$ acts on a state $|\,\mathbf{g}(c)\,\rangle$ that
in position representation is a Gaussian PDF
with zero mean and variance $c$, one obtains
\begin{equation}
\label{rule2}
\bfsog\left(a,b\right)|\mathbf{g}(c) \rangle=| \mathbf{g}(b+a^{2}c) \rangle\,.
\end{equation}

Let us now consider the piecewise
constant protocol in Eq.~(\ref{eq:protocol}). Using the Markov property, the Green function $\bfsog_{n}(t)$
for the interval $(0,t)$ is a product of the
propagators for the individual segments,
\begin{equation}
\label{Gprotocol}
\bfsog_{n}(t)=
\bfsog\left( m(t,t_{n}),v(t,t_{n})\right) \bfsog\left( m_{n-1},v_{n-1} \right)
\ldots
\bfsog\left(m_{1},v_{1}\right)\bfsog\left(m_{0},v_{0}\right)\,,
\end{equation}
where $m_j=\eta_j^{1/2}$ and $v_j=(D/\gamma_j)(1-\eta_j)$
with $\eta_j$ given in Eq.~(\ref{eq:polynomial-parameters-a}).
For the following, it is convenient to define the segment parameters
\begin{equation}
\label{eq:psidefinition}
\psi_j=\frac{\gamma_{0}}{D}v_j=\frac{\gamma_{0}}{\gamma_j}(1-\eta_j)
\end{equation}
that depend on the duration of the
segments and on the respective values of the protocol.
Notice that for any signs of the protocol parameters $\gamma_j$ the $\psi_j$ are always
positive. If the duration of the segment following the $j$-th step is
zero, its parameters are $\eta_j=1$, $\psi_j=0$. If the length tends to infinity and if $\gamma_j>0$, we have 
$\eta_j\to0$, and $\psi_j\to \gamma_0/\gamma_j$. For negative
$\gamma_j$, $\eta_j\to\infty$, and $\psi_j\to\infty$.

The initial PDF $\Phi_{\mathsf{X}}(x,0)$ is considered to
describe a particle equilibrated in the potential $V(x,0)$, i.e.\ $\Phi_{\mathsf{X}}(x,0)\propto\exp[-\beta V(x,0)]$
(assuming $k_0>0$). The initial state remains unchanged during the zeroth segment. After the
first step of the protocol, the variance either decreases (for
$\gamma_{1}>\gamma_{0}$), or increases (for
$\gamma_{1}<\gamma_{0}$). The state at the final time $t$ is
$\bfsog_{n}(t) | \mathbf{g}(v_{0}) \rangle$.  

We now use Eq.~(\ref{rule2}) to relate variances $v(t,0)$ [see Eq.~(\ref{eq:m-v-b})] at
consecutive times $t_j$ and $t_{j+1}$,
\begin{equation}
\label{eq:v-recursion}
v(t_{j+1},0)=v_0\psi_j+\eta_jv(t_j,0)\,,\hspace{2em} j=1,\ldots,n-1\,,
\end{equation}
where $v_0=v(t_1,0)=v(0,0)$. 
For the following it is convenient to introduce the scaled variances
\begin{equation}
\label{rkdefinition}
r_j=\frac{v(t_{j+1},0)}{v_0\,\eta_{j}\eta_{j-1}\ldots\eta_{1}}\,,\hspace{1em}j=1,\ldots,n-1\,,
\end{equation}
and $r_0=1$. Inserting these into Eq.~(\ref{eq:v-recursion})
yields the recursion relation
\begin{equation}
\label{eq:r-recurrence}
r_j=\phi_j+r_{j-1}
\end{equation}
with
\begin{equation}
\label{phidefinition}
\phi_j=\frac{\psi_j}{\eta_{j}\eta_{j-1}\ldots\eta_{1}}\,.
\end{equation}
After inserting $\psi_j$ from Eq.~(\ref{eq:psidefinition}) this gives Eq.~(\ref{eq:polynomial-parameters-c}).
The solution of the recursion relation~(\ref{eq:r-recurrence}) is
\begin{equation}
r_j=\phi_j+\phi_{j-1}+\ldots+\phi_1+1\,.
\end{equation}

\subsection{Propagator for the composed process}
The work done on the particle depends on the whole particle
trajectory, i.e.\ the work process $\mathsf{W}(t)$ is a
functional of the position process $\mathsf{X}(t)$. The work by itself
is not a Markov process. However, the combined process
$\left(\mathsf{X}(t),\mathsf{W}(t)\right)$ is a
time-nonhomogeneous Markov process and its joint PDF follows from the
equation \cite{Imparato/Peliti:2005, Subrt/Chvosta:2007}
\begin{equation}
\fl 
\label{FPEXW}
\frac{\partial}{\partial t}\Phi_{\mathsf{X},\mathsf{W}}(x,w,t)=
\left\{
D\frac{\partial^2 }{\partial x^2}+
\mu\frac{\partial}{\partial x}
\left[\frac{\partial}{\partial x}\,V(x,t)\right]
-\left[\frac{\partial}{\partial t}\,V(x,t)\right]\,\frac{\partial}{\partial w}
\right\}
\Phi_{\mathsf{X},\mathsf{W}}(x,w,t)\,.
\end{equation}
The first two operators on the right-hand side (RHS) are the same as in the
Fokker-Planck equation (\ref{FPEX}) and reflect the possibility that
the combined process leaves the two-dimensional interval 
$[x,x+{\rm d}x]\times[w,w+{\rm d}w]$ because of a change of the particle position.
The third operator on the RHS describes the time change of the
probabilistic weight of this interval caused by the work done on the
particle with its position fixed.  At the initial time there has been no work done yet.
Therefore the initial condition is 
$\Phi_{\mathsf{X},\mathsf{W}}(x,w,0)=\langle x | \mathbf{g}(v_{0}) \rangle \delta(w)$.

The full information given by the joint PDF can be extracted from its
double-sided Laplace transformation with respect to the work variable
$w$. We designate the transformed joint PDF as
$\tilde\Phi_{\mathsf{X},\mathsf{W}}(x,u,t)$, where $u$ is the complex
variable conjugated to $w$. After the transformation, the
partial derivative over $w$ becomes a multiplication by
$u$. We thus obtain
\begin{equation}
\label{FPEXWparabola}
\frac{\partial}{\partial t}\tilde\Phi_{\mathsf{X},\mathsf{W}}(x,u,t)=
\left\{D\frac{\partial^2 }{\partial
  x^2}+\gamma(t)\frac{\partial}{\partial x} x
+\frac{1}{2}\frac{\dot{\gamma}(t)}{D}x^{2}\frac{u}{\beta}\right\}\,
\tilde\Phi_{\mathsf{X},\mathsf{W}}(x,u,t)\,\,,
\end{equation}
where $\dot{\gamma}(t)$ is the time derivative of the protocol
function $\gamma(t)$. 

The task of solving the dynamical equation
(\ref{FPEXWparabola}) requires the calculation of a time-ordered
exponential operator. This is known to be a hard mathematical
problem. However, for the piecewise constant protocol, the time
derivative $\dot{\gamma}(t)$ in Eq.\ (\ref{FPEXWparabola}) is a sum of
delta functions. Hence the time integrals in the underlying Dyson
series can be carried out. The details of this step, in a quite
different context, are explained in \cite{Chvosta/Reineker:1999}. We
now pursue this key idea.

We insert into Eq.~(\ref{FPEXWparabola}) $\gamma(t)$ from Eq.~(\ref{eq:protocol}) and
treat the third term on the RHS as
a time-dependent perturbation with respect to the evolution controlled
by the first two terms. Using the operator notation, we
designate the Green function corresponding to
Eq.~(\ref{FPEXWparabola}) and to the protocol (\ref{eq:protocol}) as
$\bfsok_{n}(u,t)$, $t\geq t_{n}$. It is again an operator acting in
the space of $x$-functions.  After removing the multiple integrals
from the Dyson series, we find
\begin{eqnarray}
\nonumber 
\bfsok_{n}(u,t)&=&\bfsog\left(m(t,t_{n}),v(t,t_{n})\right)
\exp\left(-\frac{\gamma_{n}-\gamma_{n-1}}{2D}\,\xi\bfsox^{2}\right)\\
\nonumber
&&{}\times \bfsog\left(m_{n-1},v_{n-1}\right)
\exp\left(-\frac{\gamma_{n-1}-\gamma_{n-2}}{2D}\,\xi\bfsox^{2}\right) \\
\nonumber
&&\hspace{10em}\vdots\\
\label{averaging2}
&&{}\times\bfsog\left(m_{1},v_{1}\right)
\exp\left(-\frac{\gamma_1-\gamma_0}{2D}\,\xi\bfsox^{2}\right)
\bfsog\left(m_{0},v_{0}\right)\,,
\end{eqnarray}
where
\begin{equation}
\label{eq:xi}
\xi=\frac{u}{\beta}
\end{equation}
and $\bfsox$ is 
the position operator with matrix elements
$\langle x|\bfsox\,|x'\rangle=x\delta(x-x')$. Let us now derive a simple closed form expression for this Green function. 

Consider the action of the operator $\bfsok_{n}(u,t)$ on the
initial state $|\mathbf{g}(v_{0})\rangle$. The rightmost
operator $\bfsog\left(m_{0},v_{0}\right)$ does not alter $|\mathbf{g}(v_{0})\rangle$, i.e., 
$\bfsog\left(m_{0},v_{0}\right) |\mathbf{g}(v_{0})\rangle = |\mathbf{g}(v_{0})\rangle$. Applying the two further rightmost
operators yields
\begin{equation}
\label{Krecurrencefirststep}
\fl
\bfsog\left(m_{1},v_{1}\right)\exp\left(-\frac{\gamma_1-\gamma_0}{2D}\,\xi\bfsox^{2}\right)
|\mathbf{g}(v_{0})\rangle=
\frac{1}{\sqrt{1+\xi(\gamma_1-\gamma_0)/\gamma_0}}\,
\left|\mathbf{g}\left(v_{0}\eta_{1}R_{1}(\xi)\right)\right\rangle\,,
\end{equation}
where $R_{1}(\xi)=\phi_{1}+1/[\xi(\gamma_1-\gamma_0)/\gamma_0+1]$ and the expression on the RHS can be viewed as a non-normalized Gaussian PDF with a $\xi$-dependent complex variance. 

Repeating the procedure, there emerges a regular pattern, which allows one to write $\bfsok_{n}(u,t)$ in a compact form. This comprises a sequence of rational functions $R_j(\xi)$, $j=0,\ldots,(n-1)$, that are generated by
the recurrence relation
\begin{equation}
\label{Rkrecurrence}
R_j(\xi)=\phi_j+\frac{1}{\displaystyle \xi\epsilon_j+\frac{1}{\displaystyle R_{j-1}(\xi)}}\,\,,
\end{equation}
with $R_{0}(\xi)=1$ and $\epsilon_j$ given in Eq.~(\ref{eq:polynomial-parameters-b}).
An important property of the recurrence equation~(\ref{Rkrecurrence})
is that it coincides with
Eq.\ (\ref{eq:r-recurrence}) for $\xi=0$. Hence, $R_j(0)=r_j$ is the scaled
variance at the end of the $j$-th segment. For an arbitrary $\xi$, the
explicit solution of Eq.\ (\ref{Rkrecurrence}) is a finite regular continued fraction
\begin{equation}
\label{Rnsolution}
R_j(\xi)=\phi_j+\frac{1}{\displaystyle \xi\epsilon_j
+\frac{1}{\displaystyle \phi_{j-1}+\frac{1}{\displaystyle \xi\epsilon_{j-1}+
\frac{1}{\displaystyle\ddots\phi_{1}+\frac{1}{\displaystyle \xi\epsilon_{1}+1}}}}}\,.
\end{equation}

The final expression for the Green function is
\begin{eqnarray}
\nonumber
\fl
\bfsok_{n}(u,t)|\mathbf{g}(v_{0})\rangle
&=&\\ 
\label{Knsolution}
&&
\hspace{-4em}
\frac{1}{\displaystyle\sqrt{\prod_{j=1}^{n}[1+\xi\epsilon_jR_{j-1}(\xi)]}}
\left|\mathbf{g}\left(v(t,t_{n})+v_{0}m^{2}(t,t_{n})
\frac{R_{n-1}(\xi)\eta_{n-1}\eta_{n-2}\ldots\eta_{1}}{1+\xi\epsilon_{n}R_{n-1}(\xi)}\right)\right\rangle\!.
\end{eqnarray}
This result gives the full information on the statistics of work and position. It
allows one to calculate also correlations between any functions of these random variables.
In particular, it can be applied to generalized periodic protocols and initial conditions, which resemble
periodic processes in pumps or motors on a molecular scale \cite{Proesmans/etal:2015,Proesmans/etal:2016, Holubec/Ryabov:2017, Pietzonka/Seifert:2018, Holubec/Ryabov:2018, Solon/Horowitz:2018, Brown/Sivak:2020}. 
Some of these applications would require to extend the above findings to a time-dependent temperature, which is straightforward because it just leads to modified step parameters. A generalization of the presented methodology to other functionals of the particle position such as heat or dwelling times~\cite{Holubec2019} should be also possible.

A more difficult problem is to extend the above method to a particle initially placed at a given point, 
corresponding to a $\delta$-function. This will introduce a nonzero mean value into the Gaussian states in Eq.~(\ref{averaging2}) and significantly affect the recurrence relations. 
Mathematical structure of such generalizations can appear in various problems involving quadratic functionals of Gaussian processes, e.g., in the analysis of mean-squared displacements in single-particle tracking experiments~\cite{Grebenkov:2011, Andreanov/Grebenkov:2012}. 

\subsection{Moment generating function of work}
\label{subsec:moment-generating-function} 
For the Laplace transform of the work PDF $\Phi_n(w)$, 
the result (\ref{Knsolution}) yields
\begin{equation}
\label{Phinufinal0}
\hspace{-3em}\tilde\Phi_{n}(u)
=\left\langle{\rm e}^{-u\mathsf{W}_{n}}\right\rangle=\int_{-\infty}^{\infty}\dd x\,
\langle x|\bfsok_{n}(u,t)|\mathbf{g}(v_{0})\rangle
=\frac{1}{\displaystyle\sqrt{\prod_{j=1}^{n}[1+\xi\epsilon_jR_{j-1}(\xi)]}}\,,
\end{equation}
where $\xi=u/\beta$ [see Eq.~(\ref{eq:xi})] and the final time $t>t_n$ is irrelevant, because during the interval
$(t_n,t)$ the protocol assumes the constant value $\gamma_{n}$ and hence no work is done. 

Equation~(\ref{Phinufinal0}) is the second important result. 
However, the $u$- (or $\xi$-)dependence of the
result seems to be quite involved. For example, it is unclear how one
could calculate the derivatives of the function $\tilde\Phi_{n}(u)$. These
derivatives, evaluated at $u=0$, are needed to get the moments
$\langle\,\mathsf{W}_{n}^{k}\,\rangle$. Fortunately, the matter is simpler, because the finite product in
(\ref{Phinufinal0}) is a polynomial of degree $n$ in the complex variable $\xi$, which we prove next.

We already know that the functions $R_{k}(\xi)$ are given by the finite
continued fractions in Eq.~(\ref{Rnsolution}). These fractions can be represented as ratios of
two polynomials,
\begin{equation}
\label{PnQnration}
R_{k}(\xi)=\frac{P_{k}(\xi)}{Q_{k}(\xi)}\,\,.
\end{equation}
Upon inserting this form into the recurrence relations
(\ref{Rkrecurrence}) one arrives at a system of two linear difference
equations, which can be written in matrix form as
\begin{eqnarray}
\nonumber\hspace{-4em}
\left(\!\!\!\begin{array}{c}P_{k}(\xi)\\Q_{k}(\xi)\end{array}\!\!\!\right)&=&
\left(\!\!\!\begin{array}{cc}\xi\epsilon_{k}\phi_{k}+1&\phi_{k}\\\xi\epsilon_{k}&1\end{array}\!\!\!\right)
\left(\!\!\!\begin{array}{c}P_{k-1}(\xi)\\Q_{k-1}(\xi)\end{array}\!\!\!\right)=
\left(\!\!\!\begin{array}{cc}1&\phi_{k}\\0&1\end{array}\!\!\!\right)
\left(\!\!\!\begin{array}{cc}1&0\\\xi\epsilon_{k}&1\end{array}\!\!\!\right)
\left(\!\!\!\begin{array}{c}P_{k-1}(\xi)\\Q_{k-1}(\xi)\end{array}\!\!\!\right)\\
&=&
{\mathbb R}(\phi_{k})\,{\mathbb L}(\xi\epsilon_{k})
\left(\!\!\!\begin{array}{c}P_{k-1}(\xi)\\Q_{k-1}(\xi)\end{array}\!\!\!\right)
\label{eq:recurrence}
\end{eqnarray}
with $P_{0}(\xi)=1$ and $Q_{0}(\xi)=1$. 
In the last expression we introduced the one-parameter families of the
two-by-two matrices ${\mathbb R}(.)$ and ${\mathbb L}(.)$. They are
helpful for gaining deeper insight into the work PDF. 

The finite product in Eq.~(\ref{Phinufinal0}) equals to the polynomial $Q_{n}(\xi)$, i.e., 
\begin{equation}
\label{Phinufinal}
\tilde\Phi_{n}(u)=\frac{1}{\sqrt{Q_{n}(u/\beta)}}\,\,,
\end{equation} 
which can be readily
proved by induction: for $n=1$ the assertion is obviously true.
Suppose it is valid for
$n=k-1$. Then, using the second row in Eq.~(\ref{eq:recurrence}), it is also
valid for $n=k$. 

\section{Probability density of work}
\label{sec:work-pdf}
The expressions of the polynomials $Q_{n}(\xi)$ are made more explicit in the following two subsections~\ref{subsec:recurence_equations} and~\ref{sec:polynominal-coefficients}. The work PDF is obtained in subsection~\ref{sec:polynomial-roots} by factoring $Q_{n}(\xi)$ into its roots and by performing the inverse Laplace transform of $\tilde\Phi_{n}(u)$ via
contour integration along closed paths encircling the negative and positive real axes.

\subsection{Solution of the recurrence equations}
\label{subsec:recurence_equations}
The recurrence equations (\ref{eq:recurrence}) yield the
formal solution
\begin{eqnarray}
P_{n}(\xi)&=&\langle1,0|
{\mathbb R}(\phi_{n}){\mathbb L}(\xi\epsilon_{n}){\mathbb R}(\phi_{n-1}){\mathbb L}(\xi\epsilon_{n-1})\ldots{\mathbb R}(\phi_{1}){\mathbb L}(\xi\epsilon_{1})|1,1\rangle\,,
\label{Pnformalsolution}\\
Q_{n}(\xi)&=&\langle0,1|
{\mathbb L}(\xi\epsilon_{n}){\mathbb R}(\phi_{n-1}){\mathbb L}(\xi\epsilon_{n-1})\ldots{\mathbb R}(\phi_{1}){\mathbb L}(\xi\epsilon_{1})|1,1\rangle\,.
\label{Qnformalsolution}
\end{eqnarray}
Here and in the following we use the bra-ket notation: for a two-row
column vector with elements $x_{1},x_{2}$, we have
$x_{1}=\langle1,0|x_{1},x_{2}\rangle$, and
$x_{2}=\langle0,1|x_{1},x_{2}\rangle$. Notice that
$\langle0,1|{\mathbb R}(x)=\langle0,1|$, which has been
used in Eq.~(\ref{Qnformalsolution}). As a consequence of Eq.~(\ref{Qnformalsolution}),
the $Q_{n}(\xi)$ and hence the PDF $\Phi_{n}(w)$ depend solely on
the parameters $\phi_{k}$, $k=1,\ldots,n-1$, and  $\epsilon_{k}$, $k=1,\ldots,n$.

There exists a recurrence formula which facilitates the numerical
calculation of the polynomials $Q_{n}(\xi)$. To derive it, we start
again with the representation (\ref{Qnformalsolution}) and insert
the identity operator
$(|0,1\rangle\langle0,1|+|1,0\rangle\langle1,0|)$
just after the matrix ${\mathbb R}(\phi_{n-1})$. The result is the sum of two expressions, 
one of them being
$(1+\xi\epsilon_{n}\phi_{n-1})Q_{n-1}(\xi)$.  The other is
proportional to $\xi\epsilon_{n-1}$ and we continue with its
evaluation. After $(n-1)$ recursive steps, we obtain
\begin{equation}
\hspace{-3em}
Q_{n}(\xi)=Q_{n-1}(\xi)+\xi\epsilon_{n}
\left[
\phi_{n-1}Q_{n-1}(\xi)+\phi_{n-2}Q_{n-2}(\xi)+\ldots+\phi_{1}Q_{1}(\xi)+1
\right].
\end{equation}

In view of the following discussion, it is expedient to elaborate on the single-step and the two-steps protocol. We get 
\begin{eqnarray}
\label{P1explicit}
\rule[-1ex]{0em}{4ex}P_{1}(\xi)&=&\epsilon_{1}\phi_{1}\xi+\phi_{1}+1\,\,,
\\
\label{Q1explicit}
\rule[-1ex]{0em}{4ex}Q_{1}(\xi)&=&\epsilon_{1}\xi+1\,\,,
\\
\label{P2explicit}
\rule[-1ex]{0em}{4ex}P_{2}(\xi)&=&\epsilon_{2}\epsilon_{1}\phi_{2}\phi_{1}\xi^{2}+[\epsilon_{2}\phi_{2}(\phi_{1}+1)+\epsilon_{1}(\phi_{1}+\phi_{2})]\xi+
\phi_{2}+\phi_{1}+1\,\,,
\\
\label{Q2explicit}
\rule[-1ex]{0em}{4ex}Q_{2}(\xi)&=&\epsilon_{2}\epsilon_{1}\phi_{1}\xi^{2}+[\epsilon_{2}(\phi_{1}+1)+\epsilon_{1}]\xi+1\,\,.
\end{eqnarray}
Essentially, the form of the resulting PDF $\Phi_{1}(w)$ is given by
the root $\xi_1=-1/\epsilon_1$ of $Q_1(\xi)$, where $\epsilon_1=(\gamma_{1}-\gamma_{0})/\gamma_{0}$ [see Eq.~(\ref{eq:polynomial-parameters-b})]. If $\gamma_{1}>\gamma_{0}$ (positive step) the root is negative. Performing the inverse Laplace transform $\tilde\Phi_{1}(u)\to\Phi_{1}(w)$, we find the work PDF to vanish for negative $w$. If $\gamma_{1}<\gamma_{0}$ (negative step) the root is positive and the work PDF vanishes for positive $w$. If $\gamma_{1}$ is negative (inverted parabolic potential), the root lies in the interval $(0,1)$. The explicit result valid for any $\gamma_1$ reads
\begin{equation}
\label{Phi1w}
\hspace{-3em}
\Phi_{1}(w)=\frac{\beta}{\sqrt{\pi\epsilon_{1}\beta w}}\exp\left(-\frac{\beta w}{\epsilon_{1}}\right)
\left[\Theta(\gamma_{1}-\gamma_{0})\Theta(w)+\Theta(\gamma_{0}-\gamma_{1})\Theta(-w)\right].
\end{equation}

Simple analytical inversion $\tilde\Phi_{n}(u)\to\Phi_{n}(w)$ is also
possible in the two-step protocol. In that case the product of the two
roots of $Q_{2}(\xi)$ has the same sign as the product
$\epsilon_1\epsilon_2$. If both steps are positive
(negative), $\Phi_{2}(w)$ is nonzero for positive
(negative) $w$. If the steps differ in sign, the
work PDF is nonzero for all $w$.  For example, for two positive steps
yielding roots $-\infty<\xi_1<\xi_2<0$, we obtain
\begin{equation}
\label{Phi2w}
\Phi_{2}(w)=\beta\sqrt{\xi_1\xi_2}\,\,
{\rm I}_{0}\!\left(\frac{\beta w}{2}(\xi_2-\xi_1)\right)
\exp\left[\frac{\beta w}{2}(\xi_2+\xi_1)\right]\Theta(w),
\end{equation}
where ${\rm I}_{0}(x)$ is the modified Bessel function of zeroth order.

Figures~\ref{fig:fig2-work-distributions}(a) and (b) in Sec.~\ref{sec:recipe} show examples of the work PDF (\ref{Phi1w}) for one step, Fig.~\ref{fig:fig2-work-distributions}(c) gives an example of the work PDF in 
Eq.~(\ref{Phi2w}) for two positive steps, and in addition further examples are displayed for two and three steps.

\subsection{Coefficients of the polynomials}
\label{sec:polynominal-coefficients}
When writing the polynomials as
\begin{equation}
\label{eq:Qcoefficients}
Q_n(\xi)=\sum_{j=0}^n q_j\xi^j\,,
\end{equation}
the coefficient $q_0=1$ is fixed by the normalization and the higher-order coefficients $q_j$, $j=1,\ldots,n$ describe moments of the work $\mathsf{W}_n$, as discussed in Sec.~\ref{sec:recipe}. We now determine all $q_j$ by starting with Eq.~(\ref{Qnformalsolution}) and expressing all matrices ${\mathbb
  L}(\xi\epsilon_{k})$ as ${\mathbb I}+\xi\epsilon_{k}{\mathbb
  E}_{+}$, where ${\mathbb
  E}_{+}=|0,1\rangle\langle1,0|$. This generates
$2^{n}$ products of simpler terms.  

The coefficient $q_1$ is determined by $n$ terms which contain one matrix
${\mathbb E}_{+}$. Each of these terms factorizes into a product of two
elements. The term ending with the ket $|0,1\rangle$ is equal to
one. The term starting with $\langle1,0|$ is $r_k$. Thus
the coefficient reads
\begin{equation}
\label{qn1coefficient}
q_1=\sum_{j=1}^n\epsilon_jr_{j-1}\,\,,
\end{equation}
where the scaled variances $r_j$ are given in Eq.~(\ref{eq:r-recurrence}). 

The coefficient $q_2$ is determined by keeping the terms with products of two matrices ${\mathbb E}_{+}$. Defining
\begin{equation}
\phi_{j_{2},j_{1}}=\sum_{j=j_{1}}^{j_{2}}\phi_{j}\,,
\end{equation}
for $0\leq j_{1}<j_{2}\leq n$, the quadratic coefficient can be written as 
\begin{equation}
\label{qn2coefficient}
q_2=\sum_{j_1=1}^{n-1}\epsilon_{j_1} \phi_{j_1-1,0}
\sum_{j_2=j_1+1}^n\epsilon_{j_2}\phi_{j_2-1,j_1}\,.
\end{equation}

Continuing this reasoning, one can determine the $q_j$ for all $j=1,\ldots,n$, which are given in Eq.~(\ref{eq:qcoeefficients}). The coefficient $q_n$ appears in the factorization of $Q_n(\xi)$, see also Eq.~(\ref{PhinwLaplace}). It is given by the term in Eq.~(\ref{Qnformalsolution}) that contains $n$ matrices ${\mathbb E}_{+}$ and can also be written in the simple form
\begin{equation}
\label{qnncoefficient}
q_n=\frac{1}{\phi_{n}}\,\prod_{j=1}^{n}\epsilon_j\phi_j\,\,.
\end{equation}
All coefficients are real and continuous functions of the parameters
describing the protocol. 
Analogously, all coefficients $p_j$ defining the polynomials 
$P_n(\xi)=\sum_{j=0}^n p_j\xi^j$
with $p_0=r_n$ can be determined.

\subsection{Roots of the polynomials and explicit form of work probability density}
\label{sec:polynomial-roots}
To perform the inverse two-sided Laplace transformation of the function (\ref{Phinufinal}), the
polynomials $Q_n(\xi)$ are factored as in Eq.~(\ref{eq:q-factorization}), which leads to
the integral representation (\ref{PhinwLaplace}) of $\Phi_n(w)$. One can prove
that all roots are real, because otherwise the work PDF obtained from the
inverse Laplace transform would exhibit oscillatory behaviour and could become negative. 

\begin{figure}[t!]
\centering
\includegraphics[width = 0.8\textwidth]{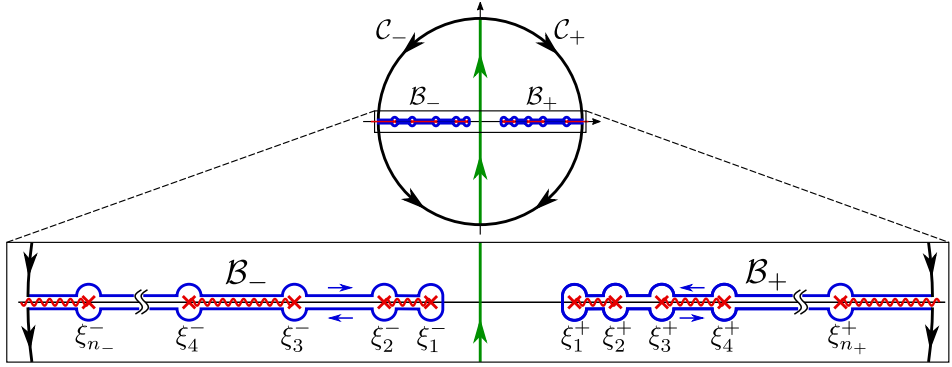}
\caption{Branch cut integration used for the inverse Laplace transform
in Eq.~(\ref{PhinwLaplace}). The integration path over the imaginary axis is replaced by the
path $\mathcal{B}_-$ for $w>0$ (corresponding to closing the contour in the left half plane) and
by the
path $\mathcal{B}_+$ for $w<0$ (right half plane).
It leads to a summation over integrals along branch cuts on the negative (positive) real axis for
positive (negative) work values. The branch cuts are indicated by the wavy lines and connect pairs $(\xi^\pm_{2j-1}, \xi^\pm_{2j})$
of roots of the polynomial $Q_n(\xi)$. If the number $n_-$ ($n_+$) of negative (positive) roots is odd, an additional integration
over the branch cut from $\xi^-_{n_-}$ ($\xi^+_{n_+}$) to $-\infty$ ($+\infty$) needs to be taken into account.}  
\label{fig:illustration-contour-integration}
\end{figure}

We now describe the details of
the contour integration leading to Eqs.~(\ref{eq:Phin-decomposition})-(\ref{eq:Cn}).
The roots $\xi_j^{\pm}$ become branch points of $\sqrt{Q_n(\xi)}$. As illustrated in
Fig.~\ref{fig:illustration-contour-integration}, we define branch cuts 
between neighbouring pairs of roots $\xi_{2j-1}^\pm$ and $\xi_{2j}^\pm$, $j=1,\ldots,m_\pm$ with $m_\pm$ given
in Eq.~(\ref{eq:mpm}), and additional branch cuts from $\xi^\pm_{n_\pm}$ to $\pm\infty$ if $n_\pm$ is odd.
If $w>0$ ($w<0$), we close the integration
path in the left (right) complex half-plane, corresponding to the
integration paths $\mathcal{C}_-$ ($\mathcal{C}_+$) 
in Fig.~\ref{fig:illustration-contour-integration}. This leads to the decomposition (\ref{eq:Phin-decomposition}) into 
$\Phi_n^\pm$ for
positive and negative work values. The integration over the circular arcs of $\mathcal{C}_\pm$ vanishes.
Each negative (positive) root yields a positive (negative) contribution to the work, i.e.\ it must correspond to a positive (negative) step of the protocol. Using Cauchy's theorem, we obtain
\begin{equation}
\label{eq:integrals-over-B}
\Phi_n^\pm(w)=-\frac{1}{2\pi i \sqrt{q_n}}\int_{\mathcal{B_\mp}}\dd\xi
\frac{\exp(\beta w \xi)}{\sqrt{\prod_{j=1}^{n_{-}} (\xi - \xi^-_j) \prod_{k=1}^{n_{+}} (\xi - \xi^+_k)}}
\end{equation}
where $\mathcal{B_-}$ 
is the path along 
the negative real axis from $-\infty$ to $\xi^-_1$, which encircles
this right-most negative root, and returns below the
negative real axis from $\xi^-_1$ to $-\infty$ (and analogous for the path $\mathcal{B_+}$
along the positive real axis).

The contribution of the branch cut between $\xi^-_{2j}$ and $\xi^-_{2j-1}$ to the integral (\ref{eq:integrals-over-B})
is
\begin{equation}
\frac{1}{\pi\sqrt{q_n}i^{n_+}(-1)^{j-1}}
\int_{\xi^-_{2j}}^{\xi^-_{2j-1}}\dd\xi
\frac{\exp(\beta w \xi)}{\sqrt{\prod_{j=1}^{n_{-}} |\xi - \xi^-_j| \prod_{k=1}^{n_{+}} |\xi - \xi^+_k|}}\,.
\end{equation}
Analogously, the branch cut between $\xi^+_{2j-1}$ and $\xi^+_{2j}$ contributes to the integral (\ref{eq:integrals-over-B}) by the term 
\begin{equation}
\label{eq:In-cut-between-positive-pairs}
\frac{1}{\pi\sqrt{q_n}i^{n_+}(-1)^{j-1}}
\int_{\xi^+_{2j-1}}^{\xi^+_{2j}}\dd\xi
\frac{\exp(\beta w \xi)}{\sqrt{\prod_{j=1}^{n_{-}} |\xi - \xi^-_j| \prod_{k=1}^{n_{+}} |\xi - \xi^+_k|}}\,.
\end{equation}
The branch cuts from $\xi^{\pm}_{n_\pm}$ to $\pm\infty$ only contribute if $n_\pm$ are odd and the integration
over them gives
\begin{equation}
\frac{\pm1}{\pi\sqrt{q_n}i^{n_+}(-1)^{m_\pm}}
\int_{\xi^\pm_{n_\pm}}^{\pm\infty}\dd\xi
\frac{\exp(\beta w \xi)}{\sqrt{\prod_{j=1}^{n_{-}} |\xi - \xi^-_j| \prod_{k=1}^{n_{+}} |\xi - \xi^+_k|}}.
\end{equation}
Summing up these contributions and considering 
that $\sqrt{q_n}=i^{n_+}\sqrt{\prod_{j=1}^{n_-} |\xi_j^-| \prod_{k=1}^{n_+} \xi_k^+}$
yields Eqs.~(\ref{eq:Phin-decomposition})-(\ref{eq:Cn}).

Some important properties of the polynomials and their roots can be deduced from general properties of the work PDF.
The normalization of $\Phi_n(w)$ is equivalent to $\tilde\Phi_n(u=0)=1$ and accordingly $Q_n(0)=1$,
see Eq.~(\ref{eq:tildephi}). This implies that there is no root at $\xi=0$. We could thus choose the
imaginary axis as integration path for the inverse Laplace transform in Eq.~(\ref{PhinwLaplace}).

If the last value of the scaled stiffness $\gamma_n$ is positive, the particle position remains confined by the parabolic potential for all times $t>t_n$. This means that the Jarzynski equality \cite{Jarzynski:1997} can be applied, which is related to the value $Q_n(1)$,
\begin{equation}
\label{jarzynski1}
\left\langle{\rm e}^{-\beta \mathsf{W}_{n}} \right\rangle=\tilde\Phi_{n}(\beta)=
\frac{1}{\sqrt{Q_{n}(1)}}=\exp\left\{-\beta\left[F_{{\rm eq}}(\gamma_{n})-F_{{\rm eq}}(\gamma_{0})\right]\right\}\,,
\end{equation}
where $F_{{\rm eq}}(\gamma)$ 
is the free energy of the equilibrium state 
in the parabolic potential with scaled stiffness $\gamma$
at temperature $T=k_{\rm\scriptscriptstyle B}/\beta$, i.e.\  $F_{{\rm eq}}(\gamma)=-[1+\log(2\pi D/\gamma)]/2\beta$.
Inserting this into Eq.~(\ref{jarzynski1}) gives
\begin{equation}
\label{eq:qn-at-one}
Q_n(1)=\frac{\gamma_n}{\gamma_0}\,.
\end{equation}
One can use the recurrence relation (\ref{eq:recurrence}) to verify this result. It is valid even for $\gamma_n\le0$ but then the average $\langle\exp(-\beta \mathsf{W}_{n})\rangle$ in the Jarzynski equality does not exist. 
From the general properties of inverse Laplace transform \cite{Doetsh:1974} (or from Eq.~(\ref{eq:In-cut-between-positive-pairs}) for $j=1$) it follows that
$\Phi_n(w)\sim |w|^{-1/2}\exp(\xi^+_1\beta w)$ for $w\to-\infty$. This implies that the average
of $\exp(-\beta \mathsf{W}_{n})$ must diverge for $\xi_1^+\in\;]0,1]$. In fact,
for $\xi_1^+=1$ the scaled stiffness $\gamma_n$ must be zero according to Eqs.~(\ref{eq:q-factorization}) and (\ref{eq:qn-at-one}). A smallest positive root $\xi_1^+\in\;]0,1[$ corresponds to a value $\gamma_n<0$. For $\gamma_n>0$, we have no roots in the interval $[0,1]$. 

\begin{figure}[t]
\includegraphics[width=\textwidth]{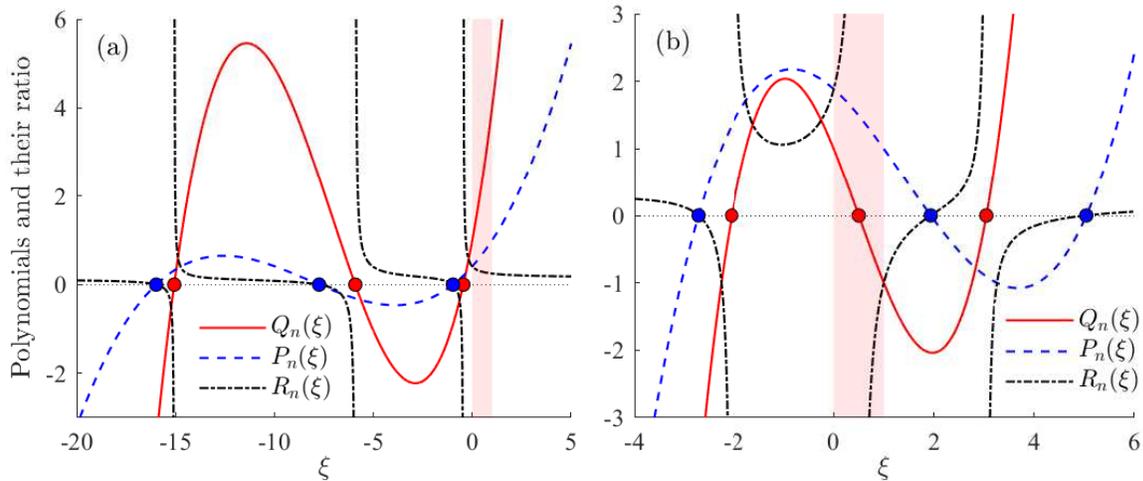}
\caption{Polynomials $P_3(\xi)$ (blue lines) and $Q_3(\xi)$ (red lines), together with their ratio $R_3(\xi)$ (black lines). Circles mark roots of the polynomials. 
In the left panel, the functions  are shown for the protocol with three positive steps: $\gamma_{0}=1$, $\gamma_1=2$, $\gamma_2=3$, and $\gamma_3=4$; durations of steps equal to 0.1 (in units of $1/\gamma_0$). All roots of $Q_3(\xi)$ are negative and hence $\Phi_3(w)$ is zero for $w<0$ as the work PDF in Fig.~\ref{fig:fig2-work-distributions}(c) plotted for a similar protocol. 
In the right panel, the functions are shown for the protocol $\gamma_{0}=1$, $\gamma_1=-1$, $\gamma_2=1$, and $\gamma_3=-1$, with step durations 0.1. Since the final value of the protocol is negative, one root occurs in the shaded region with $\xi\in\;]0,1]$.}
\label{fig:fig3-polynomials}
\end{figure}

The left panel of Fig.~\ref{fig:fig3-polynomials} demonstrates the behaviour of $Q_n(\xi)$ (red lines) for the three-step protocol where the work PDF is shown in Fig.~\ref{fig:fig2-work-distributions}(e). In this case with only positive steps all roots are negative. The right panel of Fig.~\ref{fig:fig3-polynomials} exemplifies the case of a protocol with $\gamma_n<0$, where the particle position has no equilibrium state for long times. In agreement with our general analysis above, the smallest positive root lies in the interval $]0,1[$. For completeness, we also show $P_n(\xi)$ (blue lines) and $R_n(\xi)=P_n(\xi)/Q_n(\xi)$ (black lines) in both panels.

The Crooks fluctuation theorem \cite{Crooks:1999} relates the work PDF $\overline{\Phi}_{n}(w)$ of the time-reversed protocol 
$\bar\gamma(t)$ where $\bar\gamma_j=\gamma_{n-j}$ (and durations of segments being the same) to $\Phi_{n}(w)$
as
\begin{equation}
\label{Crooks}
\overline{\Phi}_{n}(w)=\Phi_{n}(-w)\,
\exp\left\{\beta\left[w+F_{{\rm eq}}(\gamma_{n})-F_{{\rm eq}}(\gamma_{0})\right]\right\}\,.
\end{equation}
It implies the following relation between the polynomials $\overline Q_n(\xi)$ and $Q_n(\xi)$
for the reversed and original protocol:
\begin{equation}
\label{fluctuationtheoremfinal}
\overline Q_{n}(\xi)=\frac{\gamma_0}{\gamma_n}\,Q_n(1-\xi)\,.
\end{equation}
This relation can, for example, be used to check implementations of the numerical scheme discussed 
in Sec.~\ref{sec:recipe}.

\section{Summary and perspectives}
\label{sec:summary}
In summary, we studied the driven Brownian particle in a time-dependent parabolic potential and heat bath environment at temperature $T$. Initially the particle position is equilibrated and a piecewise constant time-evolution of the stiffness of the potential is considered. We derived an exact analytical result for the joint distribution of the work done on the particle and its position for a general protocol with $n$ steps. It is determined by a rational function that satisfies a recurrence relation, whose explicit solution is a finite regular continued fraction. Rewriting the continued fraction as ratio $P_n/Q_n$ of two polynomials $P_n$ and $Q_n$ of order $n$, we showed that $1/\sqrt{Q_n}$ gives the moment generating function of the work. The $Q_n$ were shown to satisfy a simple recurrence relation and roots of $Q_n$ have the following properties: (i) all roots are real and to each positive/negative step of the protocol (increasing/decreasing stiffness) corresponds one negative/positive root, (ii) the normalization condition of the work PDF requires  that there is no root at zero, (iii) if the final state of the potential has positive stiffness, the Jarzynski equality holds and requires that no root lies in the interval $[0,1]$.  With all these properties, one can very efficiently determine the work PDF for even large number $n$ of steps and thereby obtain excellent approximations for arbitrary continuous-time protocols by a suitable discretization.

From a general perspective, our results are interesting with respect to the problem of finding closed analytical formulas for a larger number of convolutions of Gaussian functions with differing parameters. This problem is important for many fields due to the ubiquitous occurrence of Gaussian processes and related quadratic approximations. In this respect it would be interesting to generalize our methodology to an arbitrary initial condition for the particle position and to a protocol, where in addition to the stiffness the position (minimum) of the parabolic potential is varying in time. From a physical perspective such generalizations may seem to be easily tractable. However, the mathematics requires a nontrivial extension of the methodology as it involves a more complex mapping between the parameters specifying the Gaussian functions. This could lead to further results for recursively defined sequences of polynomials and their relation to structured matrices, continued fractions, moment problems, and root localization \cite{Holtz/Tyaglov:2012}.  We believe that the physical treatment of corresponding problems will give valuable insight or even solutions for the mathematical convolution problem. 

A further interesting aspect concerns relations of our treatment to other studies of quadratic functionals of Gaussian processes and their PDFs. Since the work $\mathsf{W}_{n}$ is a quadratic form of correlated normal random variables $\mathsf{X}(t_{k})$, $k=0,1,\ldots,n$ (weighted by protocol steps), its PDF should be accessible numerically based on the covariance matrix of $\mathsf{X}^2(t_{k})$ as follows~\cite{Paolella:2018}: 
One finds a matrix transforming the quadratic form $\mathsf{W}_{n}$ into a weighted sum of squares of independent normal random variables. This involves computation of a square root of the covariance matrix, its inverse, and the inverse of the covariance matrix. 
The transformed sum then has a generalized chi-square distribution, whose PDF is known in a closed form. It will be interesting to see whether the algebraic structure of the polynomials $P_n$ and $Q_n$ can be used to carry out the aforementioned matrix manipulations at least semi-exactly. Another open problem is to establish an explicit connection between this algebraic structure and the eigenvalue problem for a Volterra integral equation emerging in related settings~\cite{Kac/Siegert:1947, Kac/Siegert:1947b, Slepian:1958}.  

\ack 
Financial support by the Czech Science Foundation (Project
No.\ 20-24748J) and the Deutsche Forschungsgemeinschaft (Project
No.\ 397157593) is gratefully acknowledged. We sincerely thank the
members of the DFG Research Unit FOR 2692 for fruitful discussions.
VH also thanks for support by the Humboldt foundation. 

\section*{References}
\bibliographystyle{naturemag}
\bibliography{breathing_parabola_references}
\end{document}